%% file: QSHC.tex
\documentclass[prl,aps,reprint,superscriptaddress]{revtex4-1}
\usepackage[latin1]{inputenc} 
\usepackage[T1]{fontenc} 
\usepackage[english]{babel} 
\usepackage{amsmath,amssymb,bbold} 
\usepackage{graphicx,color} 

\usepackage[breaklinks,colorlinks,linkcolor=blue,citecolor=blue,urlcolor=blue]{hyperref}

\newcommand{\eqr}[1]{Eq.~\eqref{#1}}

\newcommand{\mt}[1]{\mathrm{#1}}
\newcommand{\ket}[1]{| #1 \rangle}
\newcommand{\bra}[1]{\langle #1 |}
\newcommand{\braket}[2]{\langle #1 | #2 \rangle}

\newcommand{\im}[0]{\mathrm{i}}

\newcommand{\one}{\mathbb{1}}
\newcommand{\cc}{\mathcal{K}}

\newcommand{\UB}[0]{+} 
\newcommand{\LB}[0]{-} 
\newcommand{\cp}[0]{r} 
\newcommand{\cm}[0]{l} 

\newcommand{\LP}[0]{\mathrm{LP}} 
\newcommand{\UP}[0]{\mathrm{UP}} 

\newcommand{\dz}[0]{\hat{d}_z}
\newcommand{\ds}[0]{\hat{d}_\perp}

\begin{document}

\title{Topological Polaritons in a Quantum Spin Hall Cavity}

\author{Alexander Janot}
\affiliation{Institut f\"ur Theoretische Physik, Universit\"at Leipzig, 04009 Leipzig, Germany}
\author{Bernd Rosenow}
\affiliation{Institut f\"ur Theoretische Physik, Universit\"at Leipzig, 04009 Leipzig, Germany}
\author{Gil Refael}
\affiliation{Institute of Quantum Information and Matter, Department of Physics, California Institute of Technology, Pasadena, CA 91125, USA}
\date{\today}

%
\begin{abstract}
We study the topological structure of matter-light excitations, so called polaritons, in a  quantum spin Hall insulator coupled to photonic cavity modes.
We identify a topological invariant in the presence of time reversal (TR) symmetry, and demonstrate the existence of a TR-invariant topological phase. 
We find protected helical edge states with energies below the lower polariton branch and characteristic uncoupled excitonic states, both detectable by optical techniques. Applying a Zeeman field allows us to relate the topological index to the double coverage of the Bloch sphere by the polaritonic pseudospin.
\end{abstract}

\maketitle
%

%
Recently, topologically nontrivial states of matter with protected edge or surface states have attracted much attention~\cite{Hasan:2010,Qi:2011,Bernevig:2013}. While the first realizations were found in electronic systems~\cite{Kane:2005a,Kane:2005b,BHZ:2006,Fu:2007}, topologically nontrivial phases of periodically driven systems~\cite{Inoue:2010,Lindner:2011,Kitagawa:2010,Rechtsman:2013} and photons~\cite{Haldane:2008,Wang:2009,Hafezi:2011,Khanikaev:2013,Hafezi:2013,Susstrunk:2015} have been discovered within the last years. 
A time reversal (TR) invariant topological phase can exists when a band inversion occurs, as a function of momentum, between orbital states with different parity, and when these orbital states are coupled by spin-orbit interaction.

In electronic systems, the time reversal operator squares to minus one, $T^2=-\one$, implying the existence of degenerate Kramers pairs, such that the crossing of topological edge states is protected. 
In contrast, for bosonic systems with $T^2=+\one$, in general there is no TR invariant  topologically nontrivial phase~\cite{Schnyder:2008,Kitaev:2009,Ryu:2010}. The photonic topological insulators~\cite{Haldane:2008,Wang:2009,Hafezi:2011,Khanikaev:2013,Hafezi:2013,Susstrunk:2015} either break TR or have a built-in degeneracy, which protects edge states in a way similar to the Kramer's degeneracy in fermionic systems. 

\begin{figure}[t]
  \centering
  \includegraphics[width=0.85\hsize]{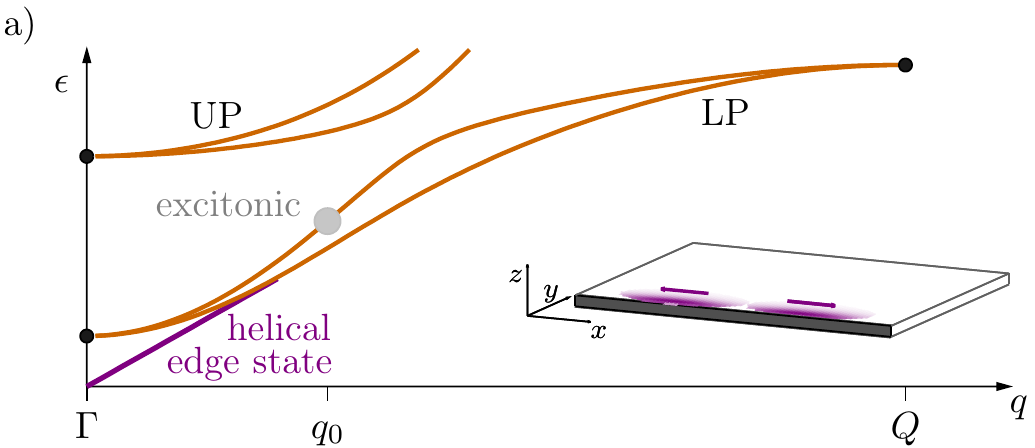}
  \includegraphics[width=0.85\hsize]{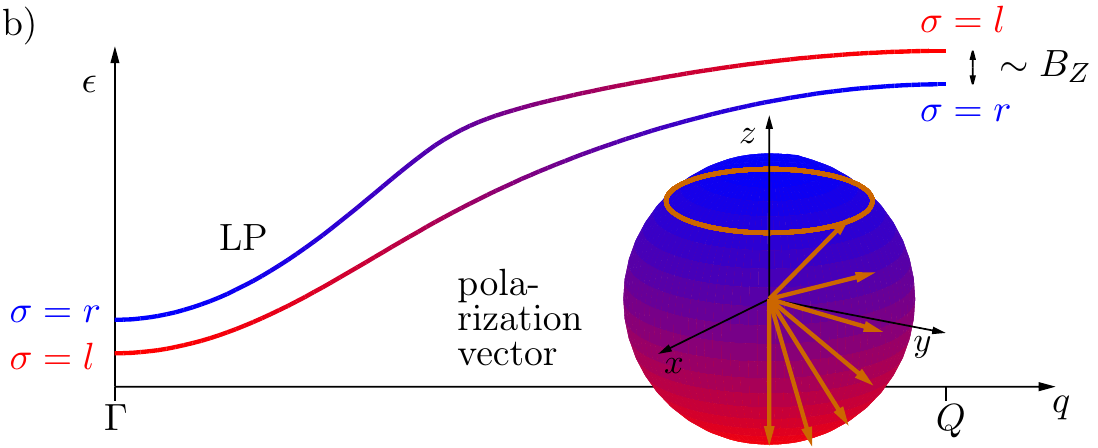}
  \caption{(color online) Defining properties of a topologically nontrivial polariton:
a) the LP and UP branches (orange lines) are split by spin-orbit coupling, except at TR invariant momenta ($\Gamma,Q$) (thick black dots). For topological polaritons purely excitonic states (gray dot) emerge along a line $\vec{q} = \vec{q}_0$, and helical edge states (thick purple line) are present below the LP.
b) with a TR breaking Zeeman field $B_Z\neq 0$, the polarization vector $\vec{n}$ of each LP-dispersion branch can be tracked over the whole Brillouin zone. The polariton is topologically nontrivial if $\vec{n}$ covers the entire Bloch sphere (inset). Here, the north (south) pole (blue (red) color) stands for right (left) circular polarization, while the $x$-$y$ plane represents linearly polarized light.
}
  \label{fig:sketch}
\end{figure}

We consider strongly coupled light-matter systems in two-dimensions, so-called polaritons, in which the bosonic polariton can inherit its topological properties from the electronic part.
Building on such an example, realized by quantum spin Hall (QSH) electrons~\cite{Kane:2005a,BHZ:2006} coupled to cavity photons, we develop a framework which allows to characterize topological states of polaritons (cf. Fig.~\ref{fig:sketch}). In contrast to recent proposals of topological polaritons~\cite{Karzig:2014,Nalitov:2014,Bardyn:2014} where TR broken systems were discussed, we focus on TR invariant topological polaritons. 
Here, we go considerably beyond previous works and i)~define a topological invariant for TR symmetric bosonic systems, ii)~explain that contrary to \cite{Schnyder:2008,Kitaev:2009,Ryu:2010} a topologically non-trivial phase is possible due to a vortex-like singularity in the exciton-photon coupling, and iii)~describe how TR-invariant topological polaritons can be detected experimentally by looking for dark excitonic states and edge states, and by studying the polarization in the presence of an external Zeeman field.

%

\emph{Main results --}
A polariton consists of an exciton coupled to a cavity photon, such that lower (LP) and upper polariton (UP) dispersion branches are formed~\cite{Deng:2010,Carusotto:2013}. The two polarization directions of the photonic component can be identified with a pseudospin~\cite{Kavokin:2004}, which can be described by an effective Hamiltonian, and whose direction can be observed by detecting photonic emission. For the LP branch it takes the form
%
\begin{equation}
	\label{eq:effectivehamiltonian}
	H_{\mt{LP}}(\vec{q}) = \epsilon_{\mt{LP}}(\vec{q}) \ + \ \vec{\sigma} \cdot  \vec{h}(\vec{q}) \ \ ,
\end{equation}
where $\vec{h}(\vec{q})$ is a momentum dependent effective magnetic field. 
With $\epsilon_{\mt{LP}}(\vec{q})$ we denote the lower polariton dispersion in the absence of pseudospin coupling, and $\vec{\sigma}$ is the vector of Pauli matrices. TR invariance $T H(\vec{q}) T^{-1} = H(-\vec{q})$ with TR operator $T = -\sigma_x \mathcal{K}$ requires that the $x$- and $y$-component of the effective magnetic field~$\vec{h}$ are even functions and the $z$-component is an odd function of momentum.
The pseudospin polarization of the eigenstate $\ket{\chi_{1,2}}$ is given by $\vec{n}_{1,2} = \bra{\chi_{1,2}} \vec{\sigma} \ket{\chi_{1,2}} = \mp \: \hat{h}$ with $\hat{h} \equiv \vec{h}/|\vec{h}|$. If $\vec{n}$  points north (south) the emitted light is right (left) polarized, while the $x$-$y$ plane represents linearly polarized light, see inset Fig.~\ref{fig:sketch}b. In the presence of an additional parity symmetry, $H_{\mt{LP}}(-\vec{q}) = H_{\mt{LP}}(\vec{q})$, one finds $h_z \equiv 0$,  and the emitted light is always linearly polarized. Physical mechanisms which lead to a non-zero effective magnetic field are, for example, a longitudinal-transverse splitting of the electromagnetic field~\cite{Kavokin:2004} or a spin-orbit coupling of the electronic building blocks as discussed below. In both cases, it turns out that $\vec{n}_{1,2}$ winds twice around the $z$-axis if $\vec{q}$ encircles the $\Gamma$-point. Due to continuity of $\vec{h}(\vec{q})$, we find that $\vec{h}(\vec{\Gamma})=0$, implying that the LP (UP) dispersion is degenerate at the $\Gamma$-point, see Fig.~\ref{fig:sketch}a;
in our QSH-polariton model, $\vec{h}(\vec{\Gamma})=0$  at all other TR invariant momenta as well.


The Chern number~\cite{TKNN:1982,Fruchart:2013} for the eigenstate $\ket{\chi_{1,2}}$ counts how many times $\vec{n}_{1,2}$ wraps around the unit sphere if $\vec{q}$ covers the Brillouin zone,
$C_{1,2} \sim \mp \int_{\vec{q}} \hat{h} \cdot (\partial_{q_x} \hat{h} \times \partial_{q_y} \hat{h})$.
Clearly, in the presence of parity symmetry, $C_{1,2} = 0$ as $h_z \equiv 0$. However, even without parity, $C_{1,2} =0$: The scalar triple product contains each component $\hat{h}_i$ with $i = x,y,z$ exactly once, and thus is an odd function of momentum, because $h_{x,y}$ is even and $h_z$ is odd. Since the integral over the Brillouin zone is invariant under momentum inversion, the Chern number has to vanish, in agreement with the general classification~\cite{Schnyder:2008,Kitaev:2009,Ryu:2010}.

For polaritons in a QSH cavity the LP eigenstates are superpositions of excitonic~$\ket{b_{1,2}}$ and photonic wave functions~$\ket{a_{1,2}}$: $\ket{\Phi_{1,2}} = \beta_{1,2} \ket{b_{1,2}} + \alpha_{1,2} \ket{a_{1,2}}$ with real coefficients $\beta_{1,2},\alpha_{1,2}$, see below for details.
An explicit calculation shows $T \ket{\Phi_{1,2}(\vec{q})} = \mp \ket{\Phi_{1,2}(-\vec{q})}$. By construction, the effective model is the projection onto the photonic sector, such that $\ket{\chi_{1,2}} = \ket{a_{1,2}}$. Both models have the same polarization vector $\vec{n}_{1,2}$, which implies that $C_{1,2} = 0$ for the microscopic model, too. 

In order to reveal the non-trivial topology of the microscopic model, we define TR~partners $\ket{\Phi_\pm} = (\ket{\Phi_2} \pm \ket{\Phi_1})/\sqrt{2}$ which transform according to $T \ket{\Phi_\pm(\vec{q})} = \ket{\Phi_\mp(-\vec{q})}$. Remarkably, the corresponding Chern numbers are non-zero, $C_\pm = \mp 2$, an interesting result given that the general classification~\cite{Schnyder:2008,Kitaev:2009,Ryu:2010} rules out non-trivial Chern numbers for eigenstates of the Hamiltonian. The robustness of $C_\pm$ is due to a vortex like non-analyticity of the exciton-photon Bloch Hamiltonian Eqs.~(\ref{eq:HP},\ref{eq:GEB}), which can only be removed when the splitting of the LP branch vanishes.
As signatures of the topological phase we find an odd number of lines of momenta~$\{\vec{q}_0\}$ encircling the $\Gamma$-point for which $\alpha_2(\{\vec{q}_0\}) = 0$, experimentally accessible by looking for a dark excitonic state at~$\{\vec{q}_0\}$. In addition, we predict polaritonic edge states. 
%
Our procedure of defining pairs of TR partners as suggested in Refs.~\cite{Roy:2009a,Roy:2009b} is analogous to defining spin-eigenstates with nontrivial spin-Chern number in a QSH insulator with broken spin-symmetry~\cite{Sheng:2006,Prodan:2009}.


If TR-symmetry is broken by a small Zeeman field, $B_Z \neq 0$, the pseudospin degeneracy at the TR invariant momenta is lifted, and two completely non-degenerate dispersion relations exist, see Fig.~\ref{fig:sketch}b. For topological polaritons with $C_\pm \neq 0$, we then find  that for each branch of the dispersion the experimentally measurable pseudospin polarization $\vec{n}(\vec{q})$ covers the Bloch sphere. Thus, in a TR broken setup, the experimentally measurable pseudospin polarization $\vec{n}(\vec{q})$ and the pseudospin model~\eqr{eq:effectivehamiltonian} allow to distinguish between topologically trivial and nontrivial polaritons. 



%
\emph{Microscopic model --}
The Hamiltonian of the two-dimensional QSH insulator in the basis of orbital states $\{ {\ket{+1/2}}$, ${\ket{+3/2}}$, ${\ket{-1/2}}$, ${\ket{-3/2}} \}$ is~\cite{BHZ:2006}
\begin{align}
	\label{eq:HBHZ}
	H_{\mt{e}}(\vec{k}) = 
	\begin{pmatrix}
		H_{\mt{e}}^\UB(\vec{k}) & 0 \\
		0 & H_{\mt{e}}^\LB(\vec{k})
	\end{pmatrix} \ ,
	\quad
	H_{\mt{e}}^\UB(\vec{k}) = \vec{d}(\vec{k}) \cdot \vec{\sigma} \ ,
\end{align}
with wavevector $\vec{k}$ and spin-orbit field $\vec{d}$.
Because of TR-symmetry $H_{\mt{e}}^\LB(\vec{k}) = {H_{\mt{e}}^\UB(-\vec{k})}^*$, where $\alpha = \{ \UB, \LB \}$ labels a pseudospin. In the following we will use the parametrization $d_{x/y} = A \sin(k_{x/y})$, $d_z = M + B (2-\cos(k_x)-\cos(k_y))$ with $A = \hbar v_F /a$, $B > 0$, and $M \in \mathbb{R}$, where $v_F$ is the Fermi velocity, $a$ the lattice spacing, and $\vec{k}$ is measured in $a^{-1}$. For $M < 0$ and $B > |M|/2$ the normalized spin-orbit field $\hat{d} \equiv \vec{d} / |\vec{d}|$ covers the Bloch sphere and the QSH insulator is topologically nontrivial~\cite{BHZ:2006}.

%
An electromagnetic field $\vec{A}$ is coupled minimally to the semi-conductor via $\vec{p} \to \vec{p} + e\vec{A}$ with elementary charge $e>0$. We work in the Coulomb gauge, linearize the Hamiltonian in $\vec{A}$ and expand the photon field in plane waves with amplitudes $\vec{A}_{\vec{q} \sigma}$ where $\vec{q}$ is the photon momentum and $\sigma$ labels the polarization~%
\footnote{
In the following, we use a basis $\vec{A}_{q,\sigma}$ with $\sigma= \{ r,l \}$ denoting circular polarization in the x-y plane, 
such that light incident under a finite angle needs to be elliptically polarized for its projection into the x-y plane to have circular polarization.
}.
We find for the optical transition matrix elements,
\begin{align}
 	\label{eq:OTM}
	g^{\sigma}_{\mu \nu}(\vec{k}',\vec{k},\vec{q}) = \delta_{\vec{k}' \vec{k}+\vec{q}} \ e\vec{A}_{q \sigma}\cdot  \bra{\psi_{\mu k'}} \  \frac{\partial H_{\mt{e}}(\vec{k})}{\partial \vec{k}} \ \ket{\psi_{\nu k}} \ ,
\end{align}
where $\mu,\nu$ label the pseudospin and band index of the eigenstates of \eqr{eq:HBHZ}.
In order to evaluate \eqr{eq:OTM} we use the Wigner-Eckart theorem~\cite{Messiah:1962} and the specific form of the basis states of \eqr{eq:HBHZ}~\cite{BHZ:2006}. We find that optical transitions do not change the pseudospin~$\alpha$.

%
A particle-hole transformation of \eqr{eq:HBHZ} yields the hole Hamiltonian $H_{\mt{h}}^\alpha(\vec{k}) = -{H_{\mt{e}}^\alpha(-\vec{k})}^*$, with wave functions $\psi^{\mt{h}}_{\alpha \, \vec{k}} = (\psi^{\mt{e}}_{\alpha \, -\vec{k}})^*$, and energies $\epsilon_{\mt{h}}(\vec{k}) = - \epsilon_{\mt{e}}(-\vec{k}) > 0$ (the chemical potential is zero).
Accounting for the attractive Coulomb interaction of electron and hole the wave function of optically active excitons is
\begin{align}
  \label{eq:fullXWF}
  \ket{\psi^{\mt{x}}_{\alpha \, \vec{q}}} = \sum_{\vec{k}} \phi_{\mt{C}}(\vec{k}) \ \ket{\psi^{\mt{h}}_{\alpha \, \vec{q}/2 - \vec{k}}} \otimes \ket{\psi^{\mt{e}}_{\alpha \, \vec{q}/2 + \vec{k}}} \ ,
\end{align}
where $\phi_{\mt{C}}(\vec{k})$ denotes the Fourier transform of the electron-hole wave function with respect to the relative coordinate. 
Semi-conductors typically have a large dielectric constant, which screens the Coulomb interaction and results in an exciton Bohr radius much larger as the lattice constant. Thus, the binding function $\phi_{\mt{C}}(\vec{k})$ is strongly peaked around $k = 0$ and $\ket{\psi^{\mt{x}}_{\alpha \, \vec{q}}} \approx \ket{\psi^{\mt{h}}_{\alpha \,\vec{q}/2}} \otimes \ket{\psi^{\mt{e}}_{\alpha \, \vec{q}/2}}$ with energy  $\epsilon_{\mt{x}}(\vec{q}) \approx \epsilon_{\mt{h}}(\vec{q}/2) + \epsilon_{\mt{e}}(\vec{q}/2)$. Such excitons are described by the Hamiltonian
\begin{align}
	\label{eq:HX}
	H_{\mt{x}}^\alpha(\vec{q}) = H_{\mt{h}}^\alpha(\vec{q}/2) \otimes \one_{\mt{e}}^\alpha + \one_{\mt{h}}^\alpha \otimes H_{\mt{e}}^\alpha(\vec{q}/2) \ ,
\end{align}
acting on a four dimensional Hilbert space of electron-hole pairs.

%
Diagonalizing \eqr{eq:HX} and projecting onto excitons gives the eigenstates $\ket{b_\alpha}$ with energy $\epsilon_{\mt{x}} = 2 |\vec{d}|$.
The cavity photon dispersion is $\omega_{\vec{q}} = \omega_0 \sqrt{1+(D/\omega_0) \: \vec{q}^{\,2}}$ with $\omega_0$ set by the cavity thickness, $D \equiv \hbar^2 c_{\mt{ph}}^2 / \omega_0 a^2$, photon velocity~$c_{\mt{ph}}$, and momentum~$\vec{q}$ measured in $a^{-1}$. Coupling excitons and right~($\cp$), left~($\cm$) circularly polarized photons yields polaritons. In the basis  $ \{ \ket{b_\UB}, \ket{b_\LB}, \ket{a_\cp}, \ket{a_\cm} \} $ the Hamiltonian takes the form
\begin{align}
  \label{eq:HP}
	H_{\mt{P}} = 
	\begin{pmatrix}
		\epsilon_{\mt{x}} \one & G \\
		G^\dagger & \omega \one
	\end{pmatrix} \ ,
\end{align}
and the exciton-photon coupling is obtained via \eqr{eq:OTM},
\begin{align}
  \label{eq:GEB}
	G = \frac{g_0}{4}
	\begin{pmatrix}
		 (1-\dz) e^{-2 \im \varphi} & -(1+\dz)\\
		 (1+\dz) & -(1-\dz) e^{2 \im \varphi}
	\end{pmatrix} 
	\ ,
\end{align}
with $\hat{d}_z = d_z / |\vec{d}|$ and $e^{\pm \im \varphi} \equiv (d_x \pm \im d_y) / |d_x \pm \im d_y|$. The coupling $g_0 \propto \epsilon_{\mt{x}}(\vec{q})$, and is proportional to the photon amplitude times a numerical constant from the transition matrix element \eqr{eq:OTM}. However, for the study of topological properties we can safely neglect the continuous $\vec{q}$-dependency and treat $g_0$ as constant.

%

\emph{Topological invariant --}  The Hamiltonian~\eqr{eq:HP} has two eigenstates for both LP and UP branch: $\ket{\Phi^\gamma_{1,2}} = b^\gamma_{1,2} \ket{b_{1,2}} + a^\gamma_{1,2} \ket{a_{1,2}}$ with $\gamma = \{ \LP ,\UP \}$, exciton $\ket{b^\gamma_{1,2}}$ and photon pseudo-spinors $\ket{a^\gamma_{1,2}}$.
We construct new basis states: $\ket{\Phi_\pm^\gamma} = (\ket{\Phi^\gamma_2} \pm \ket{\Phi^\gamma_1})/\sqrt{2}$, which are not eigenstates, but obey $T \ket{\Phi^\gamma_\pm(\vec{q})} = \ket{\Phi^\gamma_\mp(-\vec{q})}$. We define a Chern numbers  as
\begin{align}
  \label{eq:DefPolChernNumber}
	C^\gamma_\pm = -\frac{\im}{2 \pi} \, \int  \limits_{\vec{k} \: \in \: \mt{BZ}} \varepsilon_{ij} \braket{\partial_{k_i} \Phi^\gamma_\pm}{\partial_{k_j} \Phi^\gamma_\pm} \ ,
\end{align}
where $\varepsilon_{ij}$ is the Levi-Civita symbol. An explicit evaluation of \eqr{eq:DefPolChernNumber} yields $C^\LP_\pm = \mp 2$  and $C^\UP_\pm = 0$ for positive detuning, and vice versa for negative detuning. The existence of $C^\LP_\pm = \mp 2$ is due to the vortex structure on the diagonal of the coupling~\eqr{eq:GEB}. For a transition to a phase with different Chern numbers to occur, this vortex structure has to be removed, implying vanishing diagonal elements in~\eqr{eq:GEB}. At the transition point the LP-splitting vanishes and the dispersion is twofold degenerate. Hence, the Chern numbers $C^\gamma_\pm$ are protected by the splitting of LP and UP branches.

It is enlightening to analyze the photonic component
%
\begin{align}
  \label{eq:TRBph}
	\ket{a^\LP_\pm} = \frac{a^\LP_2 \mp a^\LP_1}{2} e^{\im \varphi} \ket{a_\cp} + \frac{a^\LP_2 \pm a^\LP_1}{2} e^{-\im \varphi} \ket{a_\cm} \ ,
\end{align}
of $\ket{\Phi^\LP_\pm}$.
The coefficients $a^\LP_1$, $a^\LP_2 \in \mathbb{R}$ are continuous in the entire Brillouin zone, with $a^\LP_1 \neq 0 \ \forall \: \vec{q}$, $a^\LP_2(\Gamma) = -a^\LP_1(\Gamma)$, and $a^\LP_2(Q) = a^\LP_1(Q)$, such that $a^\LP_2$ has to have an odd number of lines of zeros when going from the $\Gamma$-point to the $Q$-points. Then, the polarization vector, $\vec{n}_+ \sim \bra{a^\LP_+} \vec{\sigma} \ket{a^\LP_+}$, points north (south) at the $\Gamma$-point ($Q$-points) and winds twice around the $z$-axis in between. Thus, $\vec{n}_\pm$ covers the Bloch sphere twice. We find that the polarization vector of the excitonic component $\ket{b^\LP_\pm}$ does not cover the Bloch sphere and cannot contribute to \eqr{eq:DefPolChernNumber}.
For all wavevectors~$\vec{q}_0$ at which $a^\LP_2 = 0$, a dark exciton eigenstate, $\ket{\Phi^\LP_2} = \ket{b^\LP_2}$, exists. This constitutes a clear signature for topological polaritons.

Breaking TR-symmetry with a Zeeman field $B_Z$, we find that the LP-eigenstates $\ket{\Phi^\LP_{1,2}}$ for $B_Z\neq0$ have the same topological structure as the states $\ket{\Phi^\LP_\pm}$ for $B_Z~=~0$, i.e.~the $\ket{\Phi^\LP_\pm}$ obtained for $B_Z = 0$ can be continuously deformed into the eigenstates $\ket{\Phi^\LP_{1,2}}$ for $B_Z\neq 0$. 
Thus, if $\vec{n}_\pm$ covers the Bloch sphere in the case $B_Z=0$, then in the case $B_Z\neq0$ the polarization vector 
$\vec{n}_{1,2}$ also covers the Bloch sphere. 
%

\begin{figure}[tb]
  \centering
	\includegraphics[width=0.85\hsize]{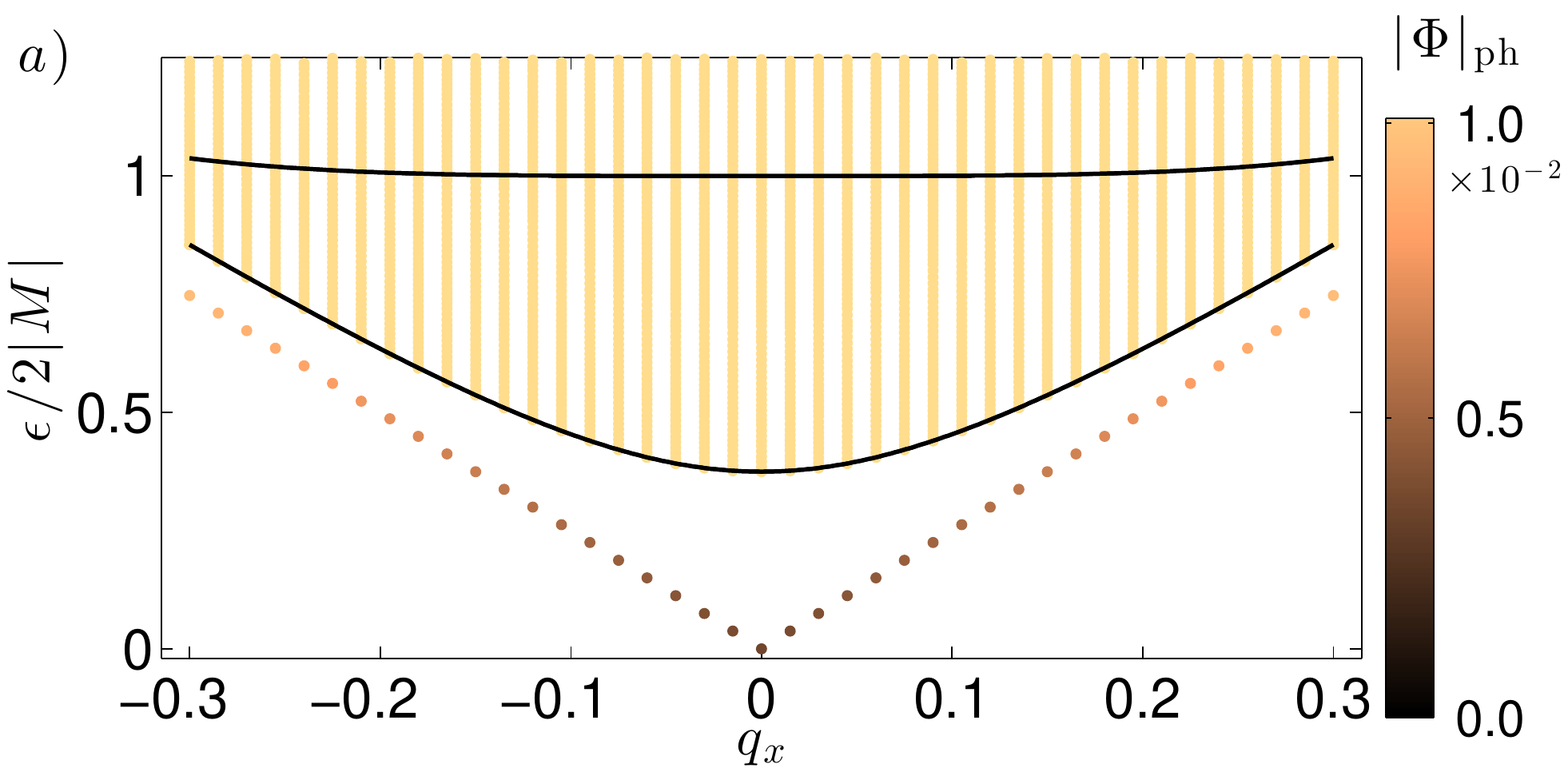} \\
	\includegraphics[width=0.85\hsize]{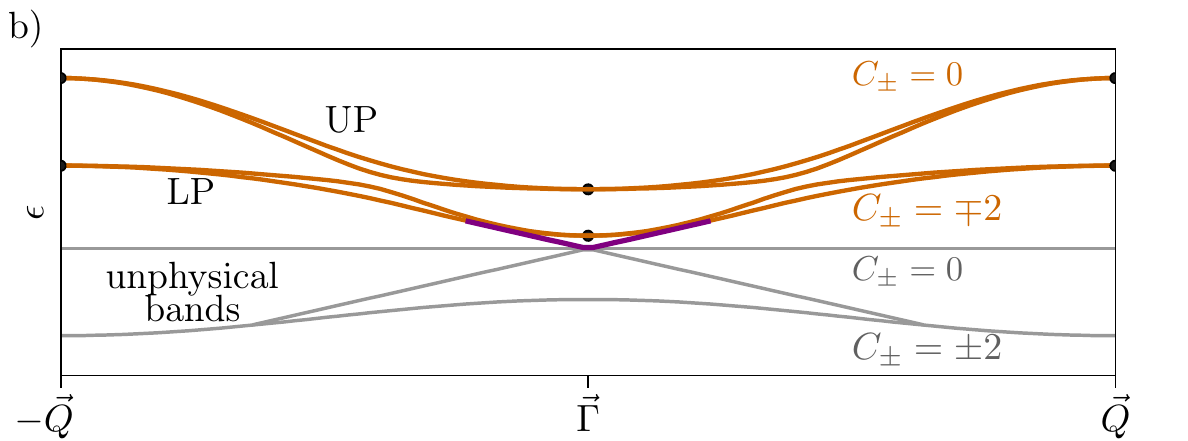}
	\caption{(color online) Spectrum of topological polaritons for a system with cylindrical geometry:
a) The colored dots depict polaritons with photonic fractions $|\Phi|_{\mt{ph}}$. The solid black lines show the LP and UP branches (LP and UP splitting not visible on this scale).
As parameters $A = 5 |M|$, $B = 25 |M|$, $\omega_0 = 3 |M|/4$, $D= 35 |M|$, $g_0 = |M|/24$, and $N = 2000$ lattice points are used.
b) The polariton branches (orange) and the unphysical electron-hole bands (light gray) are sketched schematically. Because LP-branch and  negative energy bands carry opposite Chern numbers $C_\pm$ a pair of helical edge states (the physical part is marked purple) connects both.
}
  \label{fig:SpecPolariton}
\end{figure}

%
\emph{Edge states --}
We first evaluate the electron Hamiltonian \eqr{eq:HBHZ} on a cylindrical geometry with boundaries along the $x$-direction and then couple the corresponding excitons to photon eigenmodes~\cite{SM}. The numerically obtained spectrum is shown in Fig.~\ref{fig:SpecPolariton}a. At each boundary we find one pair of edge states with energies below the LP branch.
%
How can these edge states be related to the polaritonic Chern numbers discussed above, and why extend the edge states all the way down to zero energy?
Since the excitons are a direct product of topologically nontrivial electron and hole, it is expected to find non-vanishing  Chern numbers $C^{\mt{x}}_\pm = \mp 2$. The pseudospin degeneracy of the exciton dispersion in addition to TR allows to construct an operator which commutes with the exciton Hamiltonian and behaves exactly like a fermionic TR operator, for details consult our Supp. Mat.~\cite{SM}. Due to this symmetry, the excitonic system is also in symmetry class AII and a $\mathbb{Z}_2$ index for excitons is given by $\nu = \frac{1}{4} |C^{\mt{x}}_+ - C^{\mt{x}}_-|$~\cite{Roy:2009a}. Here, the prefactor~$1/4$ was introduced because of the Chern number doubling due to the tensor product of hole and electron space~\cite{SM}.
We note that the coupled system of excitons and photons does not have any (pseudo) fermionic TR symmetry and is solely in symmetry class AI.
The non-analyticity of the polariton Hamiltonian~\eqref{eq:HP} does not allow to obtain a lattice representation and to study edge states directly. A lattice Hamiltonian is obtained by embedding the polariton space into a larger Hilbert space which contains all possible electron-hole (e-h) states~\cite{SM}, including unphysical e-h pairs which have either zero or negative energies, see Fig.~\ref{fig:SpecPolariton}b.  Since the e-h bands at negative energy  have Chern numbers~$\pm 2$, and  since the sum of all Chern numbers in the Hilbert space of polaritons and artificial e-h pairs has to be zero, the polariton Chern numbers are compensated by the negative energy e-h Chern numbers, and edge states below the LP branch emerge, cf. Fig.~\ref{fig:SpecPolariton}.

%
\emph{Effective edge model --}
The polaritonic edge states are well described by coupling excitonic edge states to photonic eigenmodes~\cite{SM}. To leading order perturbation theory we find
%
%
\begin{align}
  \label{eq:PEdgeEnergy}
  \tilde{\epsilon}(q_x) &\approx \tilde{\epsilon}_{\mt{x}}(q_x) -\left( \frac{\tilde{g}(q_x)}{\tilde{\Delta}(q_x)} \right)^2 \tilde{\Delta}(q_x) \ ,\\
  \label{eq:PEdgeWF}
  \ket{\tilde{\Phi}_{\rho \, q_x}} &\approx \ket{\tilde{\psi}^{\mt{x}}_{\rho \, q_x}} - \frac{\tilde{g}(q_x)}{\tilde{\Delta}(q_x)} \ket{\tilde{\psi}^{\mt{ph}}_{\rho \, q_x}} \ ,
\end{align}
where $\rho = \{ \mt{R}, \mt{L} \}$ labels right and left mover, $\tilde{\Delta} \equiv \tilde{\omega} - \tilde{\epsilon}_{\mt{x}}$ is the detuning between the photon $\tilde{\omega}$ and exciton energy $\tilde{\epsilon}_{\mt{x}} = \hbar v_F |q_x|$, and $\tilde{g}$ the coupling strength. The excitonic and photonic edge-state wave functions are denoted by $\ket{\tilde{\psi}^{\mt{x}}_{\rho \, q_x}}$ and $\ket{\tilde{\psi}^{\mt{ph}}_{\rho \, q_x}}$, respectively. The right (left) moving exciton carries pseudospin $\UB (\LB)$, whereas both right and left moving photon are linearly polarized with electric field parallel to the plane of incident. We find that a high photonic fraction of the edge state relies on $v_F \sim c_{\mt{ph}}$ and $g_0 \sim \omega_0$.
\begin{figure}[tb]
  \centering
  \includegraphics[width=0.48\hsize]{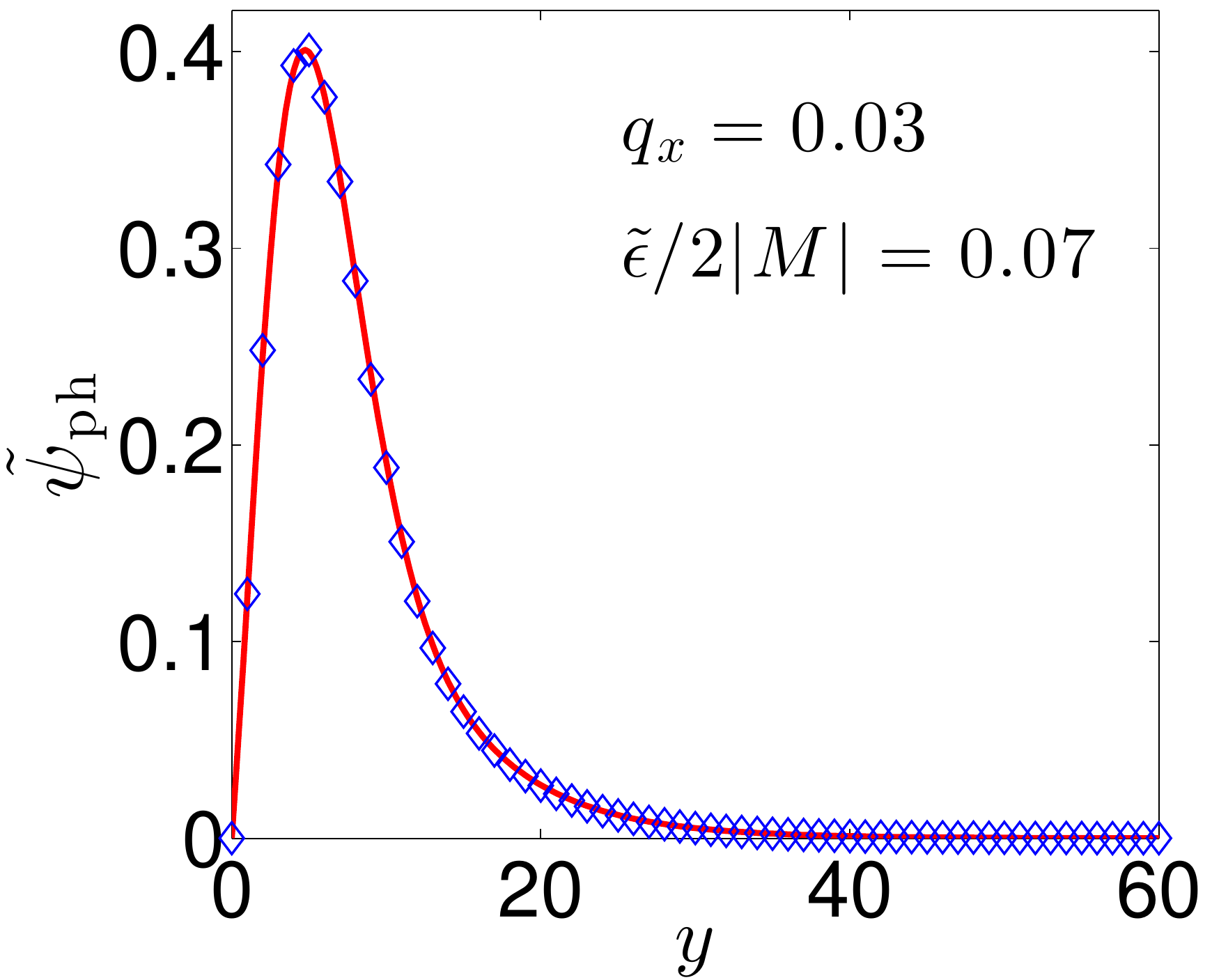} \nolinebreak \hfill
  \includegraphics[width=0.48\hsize]{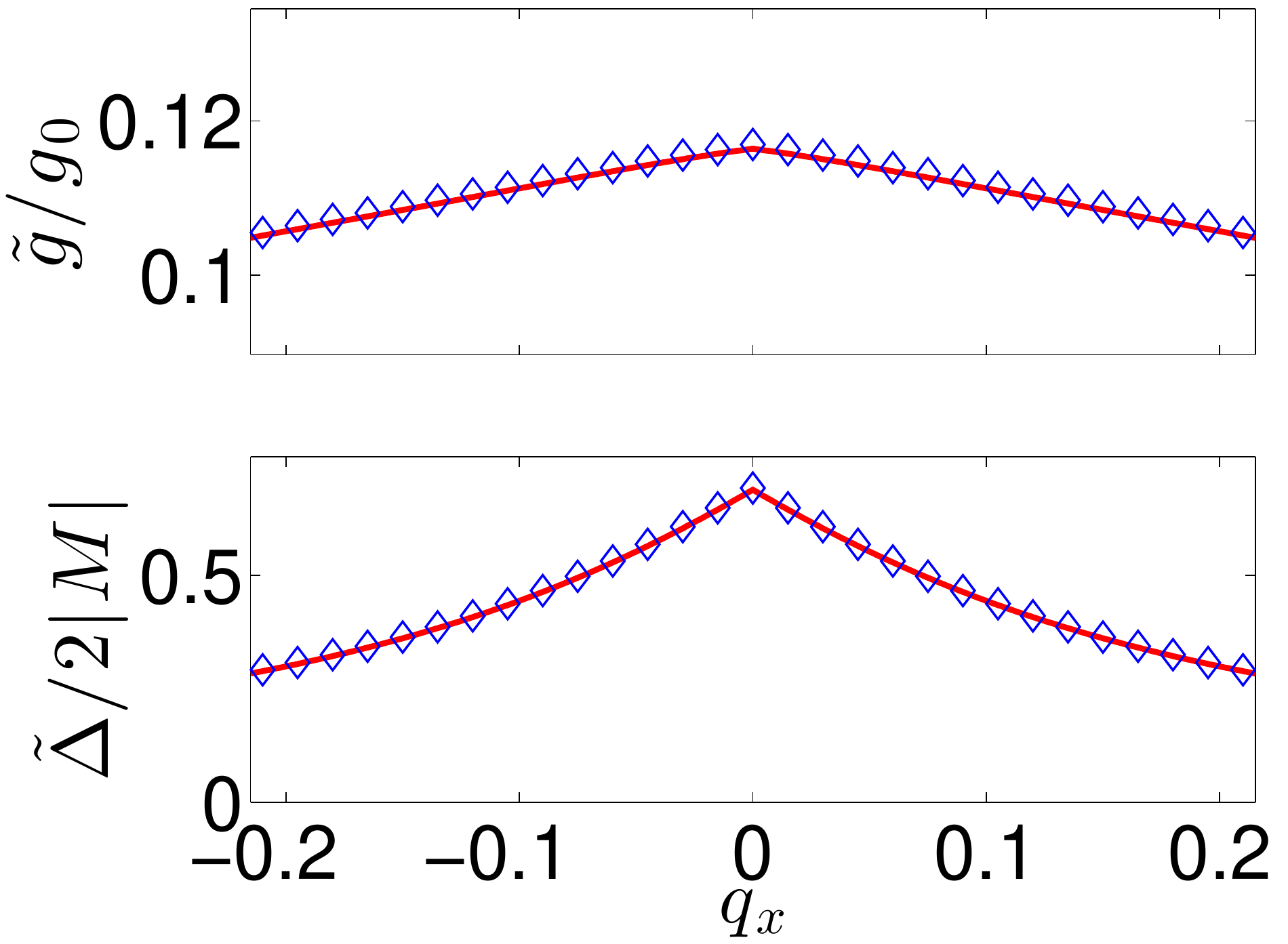}
  \caption{(color online) Comparison of analytical (lines) and numerical (symbols) results for the polaritonic edge state. Left panel: The photon wave function (cf.~\eqr{eq:PEdgeWF}) is shown.
Right panel: The coupling strength $\tilde{g}$ (upper plot) and detuning $\tilde{\Delta}$ (lower plot) of the effective model are depicted.
Same parameters as in Fig.~\ref{fig:SpecPolariton}a are used.
}
  \label{fig:Edge}
\end{figure}
For $\tilde{\epsilon}_{\mt{x}} \ll 2|M|$ our perturbative results are in very good agreement with numerics as shown in Fig.~\ref{fig:Edge}.
%
%
%
Finally, the one-dimensional effective Hamiltonian (for one edge) takes the form
\begin{align}
  \label{eq:HPEdge}
  H_{\mt{E}} = \sum \limits_{q \: \rho} \left\{ \tilde{\epsilon}_{q} \: b^\dagger_{\rho q} b_{\rho q} + \tilde{\omega}_q \: a^\dagger_{\rho q} a_{\rho q} + ( \tilde{g}_q \: b^\dagger_{\rho q} a_{\rho q} + \mt{h.c.}) \right\} \ .
\end{align}
The operator $a^\dagger_{\rho q}$ creates a right, $\rho = \mt{R}$, (left, $\rho = \mt{L}$) moving photon with momentum $q$ and dispersion $\tilde{\omega}_q$. In one dimension the elementary excitations are collective modes (plasmons) instead of excitons. Then, $b^\dagger_{\rho q}$ creates right and left moving plasmons with dispersion $\tilde{\epsilon}_{q}$. For a Luttinger liquid interactions renormalize the bare Fermi velocity~\cite{Giamarchi:2003}, and the plasmon dispersion remains linear.

%
\emph{Experimental signatures --}
The signatures of topological polaritons are dark excitonic states along a line $\{\vec{q}_0\}$ in momentum space, and edge states below the LP branch, cf. Fig.~\ref{fig:sketch}a. Both are detectable via optical techniques~\cite{Deng:2010}.
Applying a Zeeman field allows to determine the Chern number from analyzing the polarization of the bulk polaritons, see Fig.~\ref{fig:sketch}b.
Realizing topological polaritons in a nontrivial QSH insulator is challenging. However, recent developments of engineering spin-orbit coupling in polaritonic systems~\cite{Sala:2014} and accomplishing slow photons in photonic crystals~\cite{Baba:2008} may pave the way for it.

%
\emph{Conclusion --}
We have considered a TR-invariant model of polaritons in a QSH cavity, and  
introduced a topological invariant for TR partners which is stabilized by the pseudospin splitting of polaritons. In the topological phase, polaritonic edge states below the LP branch exist, as well as lines in momentum space with uncoupled excitons, both detectable via optical techniques.


We thank T.~Karzig and H.-G.~Zirnstein for discussions. 
AJ is supported by the Leipzig School of Natural Sciences BuildMoNa. BR would like to acknowledge DFG grants RO 2247/7-1 and RO 2247/8-1, and GR acknowledges NSF grant DMR 1410435, as well as the Institute of Quantum Information and Matter, an NSF center supported by the Gordon and Betty Moore Foundation.



\input{QSHC.bbl}


%
%
\clearpage
\setcounter{equation}{0}
\setcounter{figure}{0}
\renewcommand{\theequation}{S.\arabic{equation}}
\renewcommand{\thefigure}{S.\arabic{figure}}

\onecolumngrid
\section*{Supplemental Material}
%

\begin{center}
\begin{minipage}[t][][t]{0.75\hsize}
\small
In our main work we presented a symmetry argument that the Chern number of energy eigenstates has to be zero for the effective model. Here, we provide an argument that this result holds for the microscopic model, too. Furthermore, we verify that the Chern numbers of time reversed partners has to vanish for the effective model.
Then, we show how a pseudo fermionic time-reversal operator for excitons can be constructed and present the embedding of the polariton Hilbert space into an enlarged Hilbert space of artificial electron-hole pair combinations. 
Finally, we consider a system with boundaries, evaluate analytically the polaritonic edge state wave function, and present details of the effective edge-state model.
\end{minipage}
\end{center}
\vspace{4ex}
\twocolumngrid
%
%
%
\subsection*{Chern numbers}
\emph{Microscopic model --}
We found that the eigenstates of polaritons in a QSH cavity transform under time reversal (TR) according to $T \ket{\Phi(\vec{q})} = \pm \ket{\Phi(-\vec{q})}$. By symmetry arguments we will show that for any state which satisfies this behavior the corresponding Chern number defined via
\begin{align}
  \label{eqSM:DefChernNumber}
  C = - \frac{\im}{2\pi} \int \limits_{\vec{q} \, \in \, \mt{BZ}} \varepsilon_{i j} \braket{\partial_i \Phi}{\partial_j \Phi} \ ,
\end{align}
has to vanish.
We decompose the integrand into $(\braket{\partial_i \Phi(\vec{q})}{\partial_j \Phi(\vec{q})} + \braket{\partial_i \Phi(-\vec{q})}{\partial_j \Phi(-\vec{q})})/2$, which is valid because the integration over the Brillouin zone is invariant under inversion of momentum. Then, the second summand is recast: $\braket{\partial_i \Phi(-\vec{q})}{\partial_j \Phi(-\vec{q})} = \braket{\partial_i T \Phi(\vec{q})}{\partial_j T \Phi(\vec{q})} = \braket{\partial_j \Phi(\vec{q})}{\partial_i \Phi(\vec{q})}$. We used that the TR-operator and partial derivative commute, and that $T$ is anti-unitary. Now, \eqr{eqSM:DefChernNumber} takes the form $C~\sim~\varepsilon_{i j}~(\braket{\partial_i \Phi(\vec{q})}{\partial_j \Phi(\vec{q})} + \braket{\partial_j \Phi(\vec{q})}{\partial_i \Phi(\vec{q})}) = 0$.

\emph{Effective model --}
In our main work we have used symmetry arguments to show that the Chern number
\begin{align}
  	\label{eqSM:DefChernNumberPolarization}
  	C =  \frac{1}{4 \pi} \int_{\mt{BZ}} \! d^2q \; \; \vec{n} \cdot (\partial_{q_x} \vec{n} \times \partial_{q_y} \vec{n}) \ .
\end{align}
of the energy eigenstate $\ket{\chi_{1,2}}$ of the TR invariant Hamiltonian
\begin{align}
  \label{eqSM:Heffective}
  H_{\mt{LP}} = \epsilon_{\mt{LP}} + \vec{\sigma} \cdot \vec{h} \ ,
\end{align}
has to be zero. In \eqr{eqSM:DefChernNumberPolarization} the polarization vector of the eigenstate $\ket{\chi_{1,2}}$ is $\vec{n}_{1,2} = \mp \hat{h}$ with $\hat{h} = \vec{h}/|\vec{h}|$. We note that \eqr{eqSM:DefChernNumberPolarization} and \eqr{eqSM:DefChernNumber} are equivalent for $2\times2$ Hamiltonians.
For the QSH model studied in the main manuscript, $\vec{n}_{1,2}$ winds twice around the $z$-axis if $\vec{q}$ encircles the $\Gamma$-point, 
and the effective magnetic field has to vanish at the $\Gamma$-point in order to obtain a continuous Hamiltonian~\eqref{eqSM:Heffective}, which
results in an ill-defined polarization vector~$\vec{n}_{1,2}$. Thus, the Chern number~\eqr{eqSM:DefChernNumberPolarization} is formally not well defined. This raises the question of the validity of our previously presented argument.
Although the limit $\vec{q} \to \vec{\Gamma}$ is not unique, we have verified that there are no singular contributions from a neighbourhood of the $\Gamma$-point. 
Then, we can remove the $\Gamma$-point from the Brillouin zone integral, \eqr{eqSM:DefChernNumberPolarization} is applicable, and our argument for $C_{1,2} = 0$ remains valid.

The Chern numbers $C_\pm$ of the TR-partners $\ket{\chi_\pm} = (\ket{\chi_2} \pm \ket{\chi_1})/\sqrt{2}$ are defined according to \eqr{eqSM:DefChernNumberPolarization} with polarization vector $\vec{n}_{\pm} = \bra{\chi_\pm} \vec{\sigma} \ket{\chi_\pm}$. Now, we will show that $C_\pm = 0$ for the effective model~\eqref{eqSM:Heffective}. To this end we calculate explicitly $\vec{n}_\pm$.
For convenience we parametrize the effective magnetic field of \eqr{eqSM:Heffective} by $\vec{h} = (h_\perp \cos 2\varphi, h_\perp \sin 2 \varphi, h_z)^T$ with $h_\perp(-\vec{q}) = h_\perp(\vec{q})$, $\varphi(-\vec{q}) = \varphi(\vec{q}) + \pi$ and $h_z(-\vec{q}) = -h_z(\vec{q})$.
Then, the energy eigenstates take the form
\begin{align}
  	\label{eqSM:EigenstatesEffectiveModel}
  	\chi_{1,2} = \frac{1}{\sqrt{2}} 
	\begin{pmatrix}
		\phantom{\pm \; \;} \sqrt{1\mp \hat{h}_z} \; e^{-\im \varphi} \\
		\mp \sqrt{1\pm \hat{h}_z} \; e^{\im \varphi} \\
	\end{pmatrix} \ ,
\end{align}
with normalized effective magnetic field $\hat{h} = \vec{h} / |\vec{h}|$.
Using \eqr{eqSM:EigenstatesEffectiveModel}, $\vec{n}_\pm$ can be calculated in a straightforward manner,
\begin{align}
  	\label{eq:TRpolarizationvector}
  	\vec{n}_\pm = \pm \begin{pmatrix}
		- \hat{h}_z \cos 2 \varphi \\
		- \hat{h}_z \sin 2 \varphi \\
		\sqrt{1-\hat{h}_z^2}
	\end{pmatrix} \ .
\end{align}
It cannot cover the Bloch sphere, since the $z$-component is always positive (negative). Therefore, the Chern numbers of the TR-partners have to vanish, $C_\pm = 0$.

\subsection*{Pseudo fermionic TR operator for excitons}
The exciton spectrum is degenerate with respect to the pseudospin~$\alpha \in \{\UB,\LB\}$, such that $[H_{\mt{x}}, \sigma_z]=0$. 
Furthermore, $H_{\mt{x}}$ is invariant under TR. In a basis $\{\ket{b_\UB}, \ket{b_\LB}\}$ with $\sigma_z \ket{b_\pm} = \pm \ket{b_\pm}$ the excitonic TR-operator is $T = \sigma_x \cc$, which squares to one as required by the bosonic statistics of excitons. 
The product of $T$ and $\sigma_z$ commutes with $H_{\mt{x}}$ as well, and hence is a symmetry of $H_{\mt{x}}$, too. We define the anti-unitary operator $T_{\mt{F}} = \sigma_z T$. This is a fermionic TR-operator in the sense that $T_{\mt{F}} = \im \sigma_y \cc$ with $T_{\mt{F}}^2 = - \one$.

\subsection*{Enlarged polariton Hilbert space}
The BHZ Hamiltonian~\cite{BHZ:2006} in the  basis  $\{ {\ket{+1/2}}$, ${\ket{+3/2}}$, ${\ket{-1/2}}$, ${\ket{-3/2}} \}$ has the form
\begin{gather}
	\label{eqSM:HBHZ}
	H_{\mt{e}}(\vec{k}) = 
	\begin{pmatrix}
		H_{\mt{e}}^\UB(\vec{k}) & 0 \\
		0 & H_{\mt{e}}^\LB(\vec{k})
	\end{pmatrix} \ ,
\end{gather}
where $H_{\mt{e}}^\UB(\vec{k}) = \vec{d}(\vec{k}) \cdot \vec{\sigma}$ and $H_{\mt{e}}^\LB(\vec{k}) = {H_{\mt{e}}^\UB(-\vec{k})}^*$ with pseudospin $\alpha \in \{ \UB, \LB \}$, two-dimensional wavevector~$\vec{k}$ and spin-orbit field~$\vec{d}$. We found that optical transitions do not change the pseudospin~$\alpha$. Then, excitons can be characterized by a quantum number~$\alpha$, too.
In order to study the topology of these excitons we embed the exciton Hilbert space for each pseudospin~$\alpha$ into an enlarged Hilbert space; a four dimensional space spanned by the tensor-product states: $ \{ \ket{\alpha 1/2}_{\mt{h}} \otimes \ket{\alpha 1/2}_{\mt{e}}$, $\ket{\alpha 1/2}_{\mt{h}} \otimes \ket{\alpha 3/2}_{\mt{e}}$, $\ket{\alpha 3/2}_{\mt{h}} \otimes \ket{\alpha 1/2}_{\mt{e}}$, $\ket{\alpha 3/2}_{\mt{h}} \otimes \ket{\alpha 3/2}_{\mt{e}} \} $. Approximating $H^\alpha_{\mt{x}} \approx  H_{\mt{h}}^\alpha \otimes \one_{\mt{e}}^\alpha + \one_{\mt{h}}^\alpha \otimes H_{\mt{e}}^\alpha$ yields in this representation
\begin{align}
	\label{eqSM:HXUB}
	H_{\mt{x}}^\UB
	=\begin{pmatrix}
		0 & d_x - \im d_y & d_x + \im d_y & 0 \\
		d_x + \im d_y & -2d_z & 0 & d_x + \im d_y \\
		d_x - \im d_y & 0 & 2d_z & d_x - \im d_y \\
		0 & d_x - \im d_y & d_x + \im d_y & 0
	\end{pmatrix} \ .
\end{align}
Because of TR-symmetry $H_{\mt{x}}^\LB(\vec{q}) = (H_{\mt{x}}^\UB(-\vec{q}))^*$. The Hamiltonian~\eqr{eqSM:HXUB} describes an exciton state with energy $\epsilon_{\mt{x}}(\vec{q}) = 2|\vec{d}(\vec{q}/2)|$, and three unphysical pairs with holes in the conduction band and/or electrons in the valence band. The four bands of \eqr{eqSM:HXUB} have Chern numbers which result from  adding the nontrivial Chern numbers of electronic conduction and/or valence band. This results in a doubling of the exciton (hole-electron) Chern number. We find $C_{\pm} = \mp 2$ for excitons and $C_{\pm} = \pm 2$ for artificial hole-electron pairs with negative energy $-\epsilon_{\mt{x}}$. The two bands with electron and hole in the same band carry vanishing Chern numbers.

The Hamiltonian of polaritons in a QSH cavity embedded into the extended exciton space takes the form
\begin{align}
  \label{eqSM:HPSOB}
  H_{\mt{P}} = 
	\begin{pmatrix}
	H_{\mt{x}}^\UB & 0 & G_\UB \\
	0 & H_{\mt{x}}^\LB & G_\LB \\
	G_\UB^{\: \dagger} & G_\LB^{\: \dagger} & H_{\mt{ph}}
	\end{pmatrix} \ ,
\end{align}
where the exciton Hamiltonian is given in \eqr{eqSM:HXUB} and the photon Hamiltonian for right and left circularly polarized modes is of form $H_{\mt{ph}} = \omega \: \one$ with dispersion $\omega$. The coupling matrix is
\begin{align}
 	\label{eqSM:GSOBUB}
	G_\UB = \frac{g_0}{8}
	\begin{pmatrix}
		(1-\dz) \ds e^{-\im \varphi} & -(1+\dz) \ds e^{\im \varphi} \\
		(1-\dz)^2 & -\ds^{\, 2} e^{2\im \varphi} \\
		\ds^{\, 2} e^{-2 \im \varphi} & -(1+\dz)^2 \\
		(1-\dz) \ds e^{-\im \varphi} & -(1+\dz) \ds e^{\im \varphi}
	\end{pmatrix} \ ,
\end{align}
with $g_0$ being a constant, $\dz = d_z / |\vec{d}|$, and $\ds e^{\im \varphi} \equiv (d_x + \im d_y)/|\vec{d}|$. TR-symmetry demands that $G_\LB(\vec{q})~=~-G_\UB^*(-\vec{q}) \sigma_x$.
We emphasize that \eqr{eqSM:GSOBUB} is continuously defined over the entire Brillouin zone for a topologically nontrivial QSH insulator, and couples only the (physical) excitonic states to photons.

\subsection*{Polariton system on a cylindrical geometry}
First, we analyze the QSH insulator on a cylindrical geometry with periodic boundary conditions in $x$-direction and hard wall boundary conditions in $y$-direction. We compute the spectrum and eigenstates by numerical diagonalization of the lattice Hamiltonian of \eqr{eqSM:HBHZ}. The eigenfunctions are plane waves in $x$-direction with quantum numbers~$k_x$, and vanish at $y = 0,L$ with system size $L = L_y$. For given pseudospin~$\alpha$ and wavevector~$k_x$ we find $l = 1,\ldots,N$ conduction (valence) band eigenstates where $L = a N$ with $N$ lattice sites and lattice constant~$a$. These provide a basis set for electrons $\ket{\psi^{\mt{e}}_{\alpha \, k_x l}}$ and holes $\ket{\psi^{\mt{h}}_{\alpha \, k_x l}}$. For a topologically nontrivial QSH insulator the state with $l=1$ is an edge state located near a boundary. From now on we will label any edge state by a tilde.

As long as momentum was a good quantum number an approximation for the exciton wave function was given by the direct product of hole and electron wave function: $\ket{\psi^{\mt{x}}_{\alpha \, \vec{q}}} \approx \ket{\psi^{\mt{h}}_{\alpha \, \vec{q}/2}} \otimes \ket{\psi^{\mt{e}}_{\alpha \, \vec{q}/2}}$ with exciton momentum~$\vec{q}$.
This approximation fails on a cylindrical geometry, since $k_y$ is no longer a good quantum number. Nonetheless, the exciton state can be expanded in product states of electron and hole: $\ket{\psi^{\mt{x}}_{\alpha \, q_x n}} = \one_{\mt{b}} \ket{\psi^{\mt{x}}_{\alpha \, q_x n}}$ with
\begin{align}
  \label{eqSM:XWFP}
	\one_{\mt{b}} = \sum \limits_{l l'} \ket{\psi^{\mt{h}}_{\alpha \, k_x l} \ \psi^{\mt{e}}_{\alpha \, k_x l'}} \bra{\psi^{\mt{e}}_{\alpha \, k_x l'} \ \psi^{\mt{h}}_{\alpha \, k_x l}} \ ,
\end{align}
where $\ket{\psi^{\mt{e,h}}_{\alpha \, k_x l}}$ are the electron and hole eigenstates introduced above.
For sufficiently large systems we can approximate the excitonic wave function on a cylindrical geometry by projecting the excitonic eigenstates with periodic boundary conditions onto the electron-hole basis states with boundaries, namely
\begin{align}
  \label{eqSM:XWFB}
	\ket{\psi^{\mt{x}}_{\alpha \, q_x n}} &\approx \sum \limits_{l l'} c_{l l'}^{\alpha n}(q_x) \ \ket{\psi^{\mt{h}}_{\alpha \, q_x\!/2 \, l} \ \psi^{\mt{e}}_{\alpha \, q_x\!/2 \, l'}} \ , \\
	\nonumber
	c_{l l'}^{\alpha n}(q_x) &= \bra{\psi^{\mt{h}}_{\alpha \, q_x\!/2 \, l} \ \psi^{\mt{e}}_{\alpha \, q_x\!/2 \, l'}} \frac{1}{\sqrt{2}} (\ket{\psi^{\mt{x}}_{\alpha \, q_x q_n}} \pm \ket{\psi^{\mt{x}}_{\alpha \, q_x -q_n}} ) \ .
\end{align}
Above, the exciton state $\ket{\psi^{\mt{x}}_{\alpha \, q_x q_n}}$ is the plane wave solution for periodic boundary conditions with wavevector $q_{x} = 2k_x$, and $q_{n} = 2(2\pi/L)n$, where $n$ runs from $n = -N/2,\ldots,N/2-1$. We use the even superposition (plus sign) for $n \leq 0$ (cosine eigenfunction) and the odd one (minus sign) for $n>0$ (sine eigenfunctions).

We note that:
i)~The electron-hole Hilbert space is different for hard wall and periodic boundary conditions, so that $\one_{\mt{b}}$ \eqr{eqSM:XWFP} is the identity in the former and a projector in the latter space.
ii)~In the presence of edge states we replace the $n = 0$ state by the edge state wave function 
\begin{align}
  \label{eqSM:XWFE}
	\ket{\tilde{\psi}^{\mt{x}}_{\alpha \, q_x} } \approx \ket{\tilde{\psi}^{\, \mt{h}}_{\alpha \, q_x/2}} \otimes \ket{\tilde{\psi}^{\, \mt{e}}_{\alpha \, q_x/2} } \ ,
\end{align}
i.e. $c^{\alpha 0}_{l l'} = \delta_{1 l} \delta_{1 l'}$.
%
iii)~The wave function \eqr{eqSM:XWFB}
has an energy 
\begin{align}
  \label{eqSM:Xenergy}
	\epsilon_{\mt{x}}(q_x,n) = \sum_{l l'} |c^{\alpha n}_{l l'}|^2(\epsilon_{\mt{h}}(q_x/2,l)+\epsilon_{\mt{e}}(q_x/2,l')) \ ,
\end{align}
where $\epsilon_{\mt{e}}$ ($\epsilon_{\mt{h}}$) is the eigenvalue of the electron (hole) eigenfunction of the QSH insulator on a cylindrical geometry.
iv)~We note that the expansion \eqr{eqSM:XWFB} is somewhat analogous of projecting sine and cosine waves (free particles) onto a standing wave basis (particles in a box). v)~In the limit of infinite system size, wave function \eqr{eqSM:XWFB} and its spectrum \eqr{eqSM:Xenergy} converge to the solutions for periodic boundary conditions.


On a cylindrical geometry the photonic eigenmodes are
\begin{align}
  \label{eqSM:PhWF}
	\vec{A}^\sigma_{q_x m}(x,y) = \sqrt{\frac{1}{L}} e^{\im q_x x} \: \sqrt{\frac{2}{L}} \sin (q_m y) \: \vec{e}_\sigma
\end{align}
with wavevector $q_x$, $q_m = \pi/L \: m$, $m = 1,2,\ldots$, polarization vector $\vec{e}_\sigma$, and energy
\begin{align}
  \label{eqSM:Phenergy}
	\omega(q_x,q_m) = \sqrt{\omega_0^2 + \hbar^2 c_{\mt{ph}}^2 (q_x^2 + q_m^2)} \ ,
\end{align}
with $c_{\mt{ph}}$ as photon velocity and $\omega_0$ determined by the thickness of the cavity. The coupling to excitons is
\begin{align}
  \label{eqSM:coupling}
	g^{\alpha \sigma}_{n m}(q_x) = &\sum \limits_{l l'} (c_{l l'}^{\alpha n}(q_x))^* \\
		& \times \bra{\psi^{c}_{\alpha \, q_x\!/2 \, l'}} \im [\hat{H}_{\mt{e}},\hat{x}] \cdot \frac{e \vec{A}^\sigma_{q_x m}}{\hbar} \ket{\psi^{v}_{\alpha \, -q_x\!/2 \, l}} \ , \nonumber
\end{align}
where $c,v$ label the conduction and valence band, respectively.

The polaritonic modes are obtained by evaluating the coupling \eqr{eqSM:coupling} numerically and diagonalizing the corresponding polariton Hamiltonian with exciton energy \eqr{eqSM:Xenergy} and photon energy \eqr{eqSM:Phenergy} for given wavevector~$q_x$ and pseudospin~$\alpha$. We find pairs of polaritonic edge states lying energetically below the lower polariton branch.

\subsection*{Polaritonic edge state model}
%
%
Replacing $k_y \to -\im \partial_y$ in \eqr{eqSM:HBHZ} allows us to calculate analytically the edge-state wave function for hard wall boundary conditions in $y$-direction, see Ref.~\cite{Koenig:2008,Zhou:2008,Qi:2011}. For wave functions located near $y=0$~%
\footnote{
We consider a semi-infinite system with $y \in [0,L]$ and $L \to \infty$, and boundary condition $\tilde{\psi}(y=0) = 0$.
}
we find
\begin{align}
	\label{eqSM:EEWF}
	 \tilde{\psi}^{\mt{e}}_{\alpha \, k_x}(y)&= \eta_{k_x}\!(y) \, \phi^\alpha_{+1} \ ,
\end{align}
with wavevector~$k_x$, spinor~$\phi^\alpha_{\pm1}$ (eigenvector of the $\sigma_x$ Pauli matrix with eigenvalue~$\pm 1$), and real-space function
\begin{align}
  \label{eqSM:EEWF2}
	\eta_{k_x}\!(y) = 2 \frac{\sqrt{\lambda(\lambda^2-\nu^2)}}{\nu} \: e^{-\lambda y} \sinh (\nu y) \ ,
\end{align}
satisfying the boundary condition $\eta_{k_x}\!(y=0) = 0$.
The two parameters $\lambda, \nu$ (measured in inverse lattice units) are defined as
\begin{align}
  \label{eqSM:DefLambdaNu}
  \lambda  \equiv \frac{A}{B} \ , \qquad
  \nu(k_x) \equiv \sqrt{\lambda^2 - \frac{2|M|}{B} + k_x^2}
\end{align}
respectively, where $A,B,M$ are the BHZ-parameters. Above, all lengths (wavevectors) are measured in lattice units (inverse lattice units).
The edge state exists if $|k_x| < \sqrt{2|M|/B}$, and has an energy
\begin{align}
  \label{eq:XEdgeDispersion}
  {\tilde{\epsilon}_{\mt{e}}^{\, \alpha}}(k_x) = \alpha \: \hbar v_F k_x \ .
\end{align}
Edge states located near $y=L$ have $\phi^\alpha_{-1}$ spinors and energies ${\tilde{\epsilon}_{\mt{e}}^\alpha}(k_x) = -\alpha \hbar v_F k_x$. In the following we will focus on the boundary $y=0$.

%
Now, the coupling \eqr{eqSM:coupling} is evaluated using the approximation \eqr{eqSM:XWFE} and the result \eqr{eqSM:EEWF}. We find that only p-polarized light (linearly polarized light with electric field parallel to the plane of incident which is perpendicular to the $y$-direction) couples: $g^{\alpha \sigma = p}_m \equiv g^{\alpha}_m \neq 0$, whereas s-polarized light (electric field perpendicular to the boundary) does not: $g^{\alpha \sigma = s}_m = 0$.
We find
\begin{align}
  \label{eqSM:Ecoupling}
	g^{\pm}_m(k_x) = -\frac{\im}{\sqrt{2}} \frac{g_0}{2} \sqrt{\frac{2}{L}} \int_y {\eta_{k_x}\!(y)}^2 \sin (\frac{\pi m}{L} y) \ ,
\end{align}
for both pseudospins $\alpha = \pm$.

%
%
As long as the exciton bulk states are energetically much higher as the edge state, $\tilde{\epsilon}_{\mt{x}} = \hbar v_F |q_x| \ll 2|M|$, we can neglect those and use an effective description,
\begin{align}
  \label{eqSM:HPE1}
  H_{\mt{E}}'(q_x) =
	\begin{pmatrix}
		\tilde{\epsilon}_{\mt{x}}(q_x) & 0 & \underline{g}^\UB(q_x) \\
		0 & \tilde{\epsilon}_{\mt{x}}(q_x) & \underline{g}^\LB(q_x) \\
		(\underline{g}^{\UB}(q_x))^\dagger & (\underline{g}^{\LB}(q_x))^\dagger & \underline{\omega}(q_x)
	\end{pmatrix} \ ,
\end{align}
where $\underline{g}^\alpha$ is a row vector in the photon-space, see \eqr{eqSM:Ecoupling}, and  $\underline{\omega}(q_x) = \omega(q_x,q_m) \delta_{m m'}$ a diagonal matrix, see \eqr{eqSM:Phenergy}.
In leading order degenerate perturbation theory we find
\begin{align}
  \nonumber
  \tilde{\epsilon}(q_x) &\approx \tilde{\epsilon}_{(0)}(q_x) + \tilde{\epsilon}_{(2)}(q_x) \\
	&= {\tilde{\epsilon}_{\mt{x}}}(q_x) - \sum \limits_{m} \frac{|g^\pm_{m}(q_x/2)|^2}{\Delta_{m}(q_x)} \ ,
  \label{eqSM:PEEnergy1} \\
  \nonumber
  \ket{\tilde{\Phi}_{\rho \,q_x}} &\approx  \ket{\tilde{\Phi}_{\rho \, q_x}^{(0)}} + \ket{\tilde{\Phi}_{\rho \, q_x}^{(1)}} \\ 
	&= \ket{\tilde{\psi}^{\mt{x}}_{\rho \, q_x}} - \sum \limits_{m} \frac{({g}^{\pm}_{m}(q_x/2))^*}{\Delta_{m}(q_x)} \ket{\vec{A}^{\sigma=p}_{q_x m}} \ .
  \label{eqSM:PEWF1}
\end{align}
Above, we have defined the detuning $\Delta_{m}(q_x) \equiv \omega(q_x,q_m) - {\tilde{\epsilon}_{\mt{x}}}(q_x)$, and an index $\rho$ which labels right ($\mt{R}$) and left ($\mt{L}$) moving edge states. The excitonic component of the right (left) mover has pseudospin $\alpha = \UB (\LB)$, whereas the photonic component is p-polarized independently of the direction of motion. Since $g_m^\pm$ is equal for both pseudospins, the edge-state energy is degenerate for $\rho = \{ \mt{R},\mt{L} \}$.
Evaluating the photonic part of $\tilde{\Phi}_{\rho q_x}(y)$ \eqr{eqSM:PEWF1} results in a wave function which is exponentially localized near the boundary, too.
We define the photon fraction as $F \equiv \sqrt{ | P_{\mt{ph}} \; \tilde{\Phi}_{\rho \, q_x}|^2 }$ where $P_{\mt{ph}}$ projects onto the photonic part, so that \eqr{eqSM:PEWF1} yields
\begin{align}
  \label{eqSM:PhotonFraction}
	F(q_x) \approx \sqrt{\sum \limits_{m} \frac{|{g}^\pm_{m}(q_x/2)|^2}{{\Delta_{m}(q_x)}^2}} \ .
\end{align}

A simple effective model for right or left moving polaritonic edge states is
\begin{align}
  \label{eqSM:HP_XEdgePhEdge}
  H_{\mt{E}}(q_x) =
	\begin{pmatrix}
		{\tilde{\epsilon}_{\mt{x}}}(q_x) & \tilde{g}(q_x) \\
		(\tilde{g}(q_x))^* & \tilde{\omega}(q_x)
	\end{pmatrix} \ ,
\end{align}
where $\tilde{\epsilon}_{\mt{x}}(q_x) = \hbar v_F |q_x|$ is the exciton energy, $\tilde{\omega}$ the energy of the localized photon wave function and $\tilde{g}$ an effective coupling. Diagonalizing \eqr{eqSM:HP_XEdgePhEdge} and expanding in powers of $\tilde{g}/\tilde{\Delta}$ yields
\begin{align}
  \label{eq:PEEnergy2}
  \tilde{\epsilon}(q_x) &\approx \tilde{\epsilon}_{\mt{x}}(q_x) -\left( \frac{\tilde{g}(q_x)}{\tilde{\Delta}(q_x)} \right)^2 \tilde{\Delta}(q_x) \ ,\\
  \label{eq:PEWF2}
  \ket{\tilde{\Phi}_{\rho \, q_x}} &\approx \ket{\tilde{\psi}^{\mt{x}}_{\rho \, q_x}} - \frac{\tilde{g}(q_x)}{\tilde{\Delta}(q_x)} \ket{\tilde{\psi}^{\mt{ph}}_{\rho \, q_x}} \ ,
\end{align}
with detuning $\tilde{\Delta} \equiv \tilde{\omega} - \tilde{\epsilon}_{\mt{x}}$. This allows us to extract
\begin{align}
  \label{eqSM:ECoupling}
	\tilde{g}(q_x) &= \frac{1}{F(q_x)} \sum \limits_{m} \frac{|g^\pm_{m}(q_x/2)|^2}{\Delta_{m}(q_x)} \ , \\
	\label{eqSM:PhEDispersion}
	\tilde{\Delta}(q_x) &= \frac{1}{F(q_x)^2} \sum \limits_{m} \frac{|g^\pm_{m}(q_x/2)|^2}{\Delta_{m}(q_x)} \ , \\
	\label{eqSM:PhEWF}
	\ket{\tilde{\psi}^{\mt{ph}}_{\rho \, q_x}} &= \frac{1}{F(q_x)} \sum \limits_{m} \frac{({g}^{\pm}_{m}(q_x/2))^*}{\Delta_{m}(q_x)} \ket{\vec{A}^{\sigma=p}_{q_x m}} \ ,
\end{align}
by comparing with \eqr{eqSM:PEEnergy1} and \eqr{eqSM:PEWF1}.

\newpage

\end{document}

%% file: QSHC.bbl
%